\documentclass{aa}  
\usepackage{graphicx}
\usepackage{txfonts}
\begin{document}

\title{First detection of a disk free of volatile elements around a young A-type star: A sign of collisions between rocky planets?\thanks{Based on observations collected at the European Southern Observatory under ESO programme 0101.C-0902(A).}}
\titlerunning{First detection of a disk free of volatile elements around a young A-type star}
   \author{M. E. van den Ancker\inst{1}
          \and
          N. P. Gentile Fusillo\inst{1}
          \and
          T. J. Haworth\inst{2}
          \and
          C. F. Manara\inst{1}
          \and
          P. A. Miles-P\'aez\inst{1}
          \and 
          R. D. Oudmaijer\inst{3}
          \and 
          O. Pani\'c\inst{3}
          \and 
          D. J. M. Petit dit de la Roche\inst{1}
          \and 
          M. G. Petr-Gotzens\inst{1}
          \and
          M. Vioque\inst{4,5}
          }

   \institute{European Southern Observatory, Karl-Schwarzschild-Str. 2, D--85748 Garching bei M\"unchen, Germany\\
             \email{mvandena@eso.org}
              \and
              Astronomy Unit, School of Physics and Astronomy, Queen Mary University of London, London E1 4NS, UK
            \and
              School of Physics and Astronomy, University of Leeds, Woodhouse Lane, Leeds, LS2 9JT, UK
            \and
              Joint ALMA Observatory, Alonso de C\'ordova 3107, Vitacura 763-0355, Santiago, Chile
            \and
              National Radio Astronomy Observatory, 520 Edgemont Road, Charlottesville, VA 22903, USA
            }

   \date{Received; accepted }

% \abstract{}{}{}{}{} 
 
  \abstract
   {}
   {We present the first detailed analysis of the astrophysical parameters of the poorly studied Sco-Cen member HD\,152384 and its circumstellar environment.}
   {We analyse newly obtained optical-near-IR XSHOOTER spectra, as well as archival TESS data, of HD\,152384. In addition, we use literature photometric data to construct a detailed spectral energy distribution (SED) of the star.}
   {The photospheric absorption lines in the spectrum of HD\,152384 are characteristic of a A0\,V star, for which we derive a stellar mass of 2.1 $\pm$ 0.1 M$_\odot$ and a stellar age $>$ 4.5 Myr. Superimposed on the photospheric absorption, the optical spectrum also displays double-peaked emission lines of Ca\,{\sc ii}, Fe\,{\sc i}, Mg\,{\sc i} and Si\,{\sc i}, typical of circumstellar disks. Notably, all Hydrogen and Helium lines appear strictly in absorption. A toy model shows that the observed emission line profiles can be reproduced by emission from a compact (radius $<$ 0.3 au) disk seen at an inclination of $\sim24^\circ$. Further evidence for the presence of circumstellar material comes from the detection of a moderate infrared excess in the SED, similar to those found in extreme debris disk systems.}
   {We conclude that HD\,152384 is surrounded by a tenuous circumstellar disk which, although rich in refractory elements, is highly depleted of volatile elements. To the best of our knowledge such a disk is unique within the group of young stars. However, it is reminiscent of the disks seen in some white dwarfs, which have been attributed to the disruption of rocky planets. We suggest that the disk around HD\,152384 may have a similar origin and may be due to collisions in a newly formed planetary system.}

   \keywords{circumstellar matter -- protoplanetary disks -- stars: emission-line -- stars: pre-main sequence -- 
stars: HD 152384}
   \maketitle
%
%-------------------------------------------------------------------

\section{Introduction}
HD\,152384 (HIP\,82714, $G$ = 7.01) is a poorly studied early-type star located in the constellation of Scorpius \citep{Houck1982, McDonald2012}. From its parallax and proper motions, \citet{Rizzuto2011} and \citet{Galli2018} identified HD\,152384 as a member of the 5--10 Myr Sco-OB2 association, suggesting a young age. Based on an anomaly between the Hipparcos and Gaia DR2 proper motions, \citet{Kervella2019} classified HD\,152384 as a suspected binary system. More recently, the multiple nature of HD\,152384 was confirmed by \citet{El-Badry2021}, who identified the source Gaia EDR3 6026288713470340608 ($G$ = 13.58, separation 77.20$\arcsec$ or 9908 au) as its companion.

HD\,152384 is included in a new sample of 246 Gaia-selected young intermediate-mass stars for which we are currently analysing UV-near-IR spectroscopy obtained with the XSHOOTER instrument on ESO's Very Large Telescope (VLT). Within this sample, and to our knowledge within the entire class of YSOs, HD\,152384 shows a unique optical-near-IR spectrum, on which we report in this {\it Letter}. 

%--------------------------------------------------------------------
\section{Observations}
\begin{figure*}
   \centering
   \includegraphics[angle=-90, width=13cm]{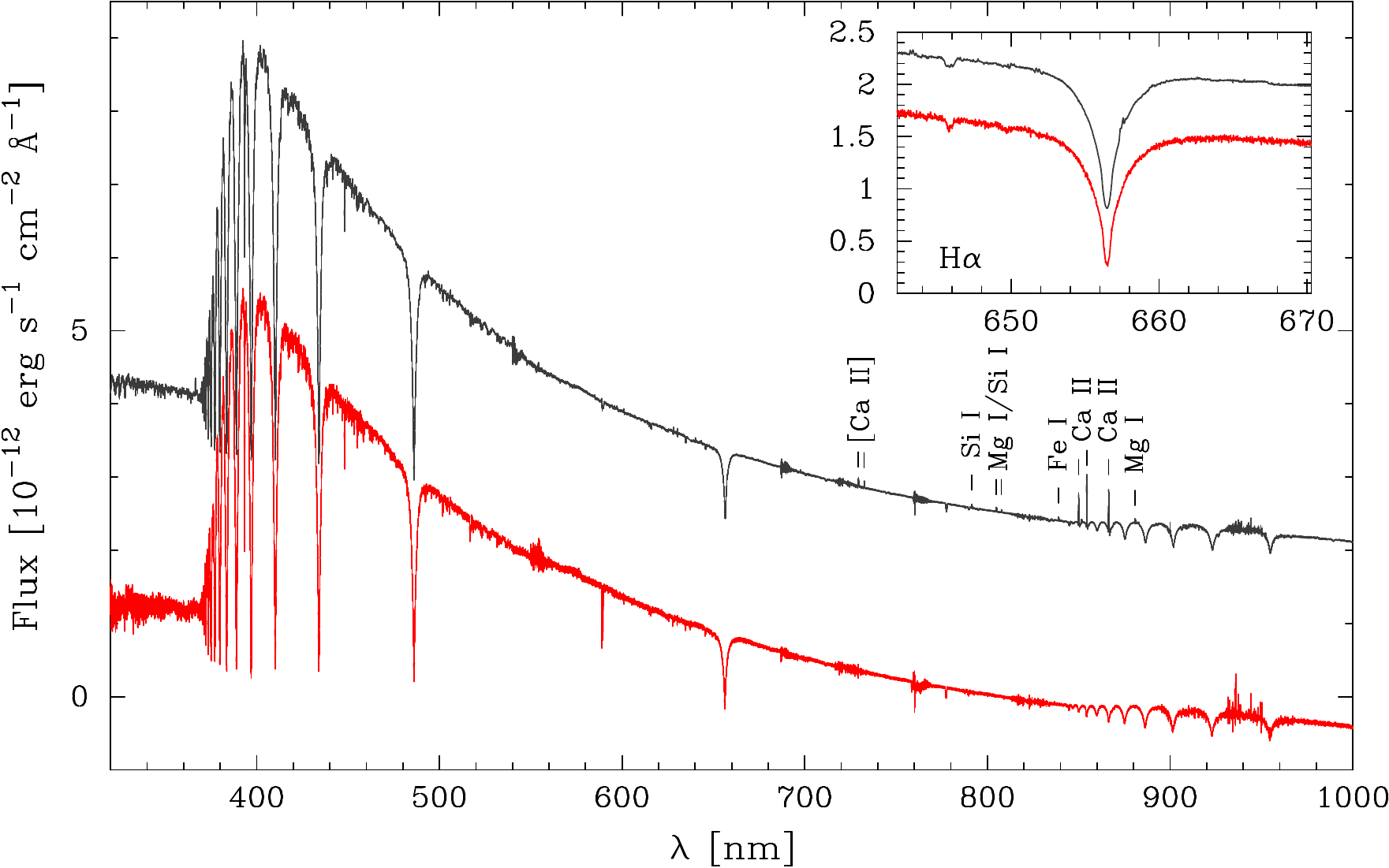}
   \caption{UV-Optical XSHOOTER spectrum of HD\,152384 (black line) with the location of the most prominent emission lines indicated. Also shown (red line) is the spectrum of HD\,207222 (Sp. Type = A0), obtained from the XSHOOTER Spectral Library \citep{Gonneau2020}, shifted for clarity. The inset shows the region around the H$\alpha$ line (656.3 nm).}
              \label{Fig1}
    \end{figure*}

A high-quality spectrum of HD\,152384 was obtained with XSHOOTER -- a wide spectral range (300--2400 nm) spectrograph on the VLT \citep{Vernet2011} -- on July 2, 2018. Total integration times were 60 seconds, 60 seconds and 100 seconds for XSHOOTER's UV, visual and near-IR arms, respectively. The XSHOOTER entrance slits were oriented along the parallactic angle ($-93.9^\circ$) to minimise atmospheric diffraction losses and slit widths were set to 1.0", 0.9" and 0.6" for the UV, visual and near-IR arms, resulting in spectra with a spectral resolution of 5400, 8900 and 8100 for these three arms. Data were reduced using the XSHOOTER data reduction pipeline version 3.3.5\footnote{http://www.eso.org/sci/software/pipelines/}, after which the spectrum was corrected for telluric absorption using a synthetic transmission spectrum computed using the molecfit package\footnote{http://www.eso.org/sci/software/pipelines/skytools/molecfit}, fitted to the visual and near-IR data. The resulting spectrum is shown in Fig.~\ref{Fig1}.

%-------------------------------------- Two column figure (place early!)

\section{Stellar Parameters}
\begin{table*}
  \caption[]{Stellar properties of HD\,152384 derived in this paper.}
  \label{Table1}
\tabcolsep0.16cm
\vspace*{-0.4cm}
\begin{flushleft}
\begin{tabular}{ccccccccccc}
\hline\noalign{\smallskip}
$d$        & Sp. Type & $T_{\rm eff}$ & $v_{\rm rad}$   & $v \sin i$  & $L_\star$ & $R_\star$ & $M_\star$ & $P_{\rm rot}$  & $i$  & Age\\
$[$pc] & & [K] & [km~s$^{-1}$]  & [km~s$^{-1}$] & [L$_\odot$] & [R$_\odot$] & [M$_\odot$] & [hours] & [$^\circ$] & [Myr]\\
\noalign{\smallskip}
\hline\noalign{\smallskip}
128.3 $\pm$ 0.6 & A0\,V   & 9500 $\pm$ 200 & $-10$ $\pm$ 2 & 116 $\pm$ 20 & 27.4 $\pm$ 2.1 & 1.9 $\pm$ 0.2 & 2.1 $\pm$ 0.1 & 8.286 $\pm$ 0.006 & 24 $\pm$ 4 & > 4.5\\
\noalign{\smallskip}
\hline
\end{tabular}
\end{flushleft}
\end{table*}

   \begin{figure}
   \centering
   \includegraphics[angle=-90, width=\hsize]{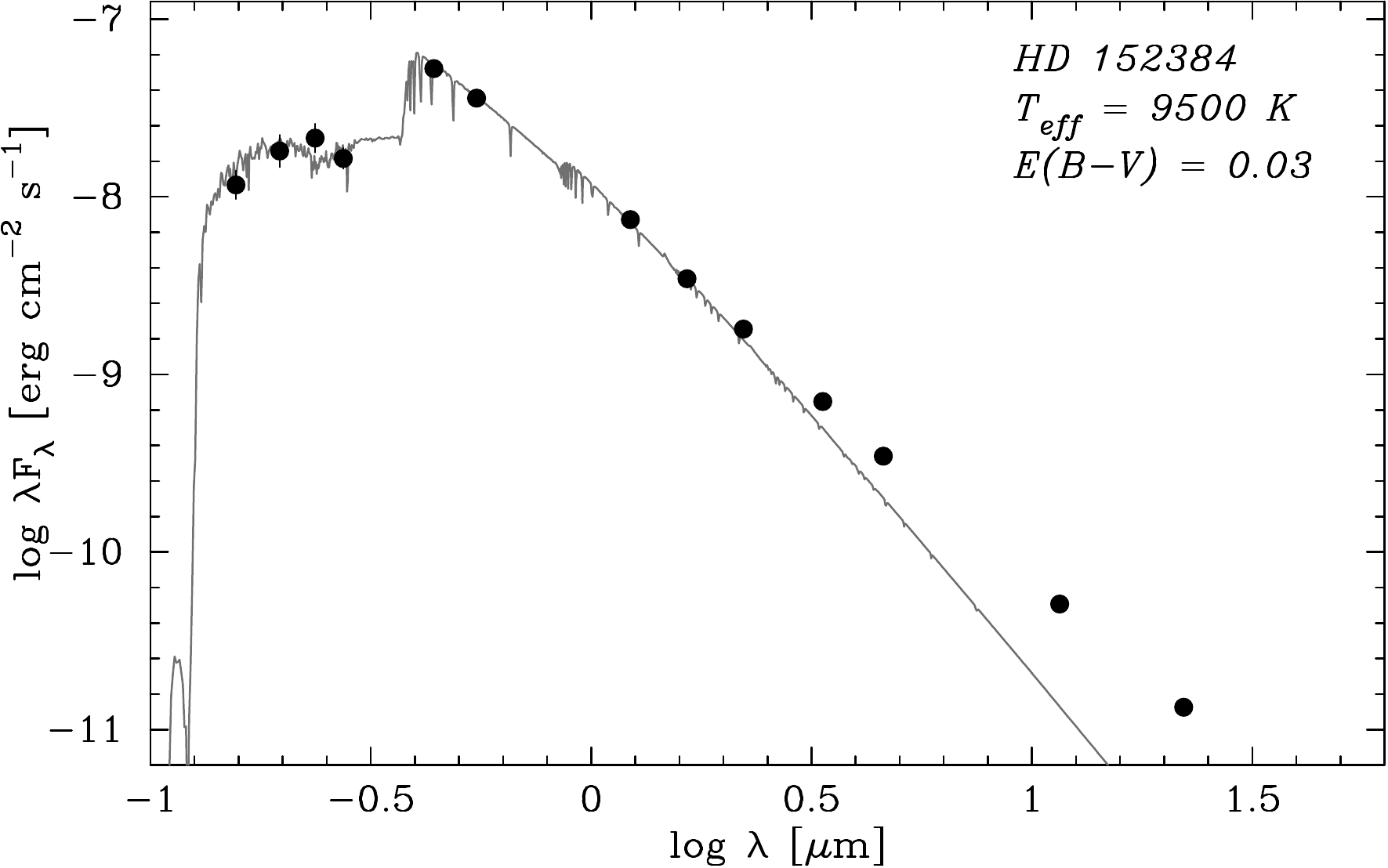}
   \caption{Spectral Energy Distribution of HD\,152384. Observed photometry (black dots) is compared to a \cite{Kurucz1991} model for a A0 ($T_{\rm eff}$ = 9500 K) star (grey line), reddened using the \cite{Cardelli1989} normal extinction law and a colour excess E(B-V) = 0\fm03.}
              \label{Fig2}
    \end{figure}

   \begin{figure}
   \centering
   \includegraphics[width=\hsize]{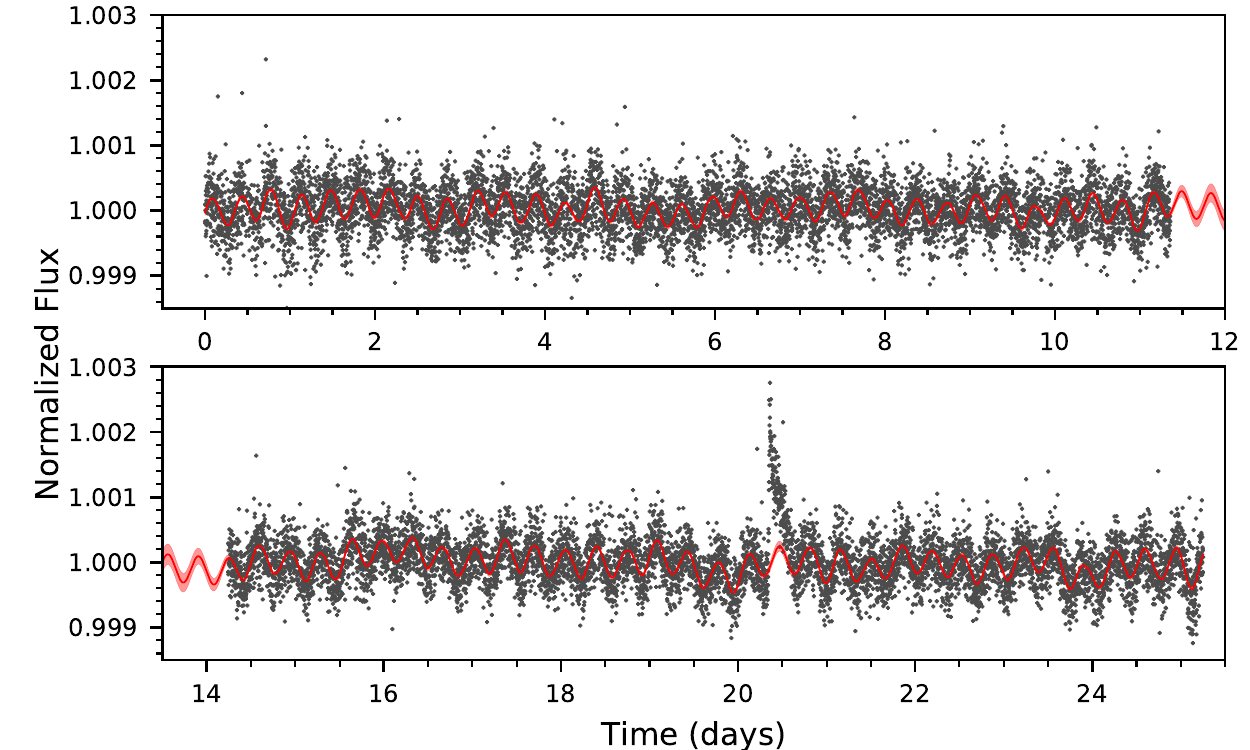}
   \caption{TESS broad-band (600--1000 nm) normalized photometric time series of HD\,152384 (black dots), covering the time period May 21, 2019 to June 18, 2019. In addition to the periodic variations on a 8.286 hour time scale, a flare can be seen in the data around 20.5 days. The red line shows the Gaussian Process fitted to the data.}
              \label{Fig3}
    \end{figure}
   %
%
%-------------------------------------- Two column figure (place early!)
   \begin{figure}
   \centering
   \includegraphics[width=8.5cm]{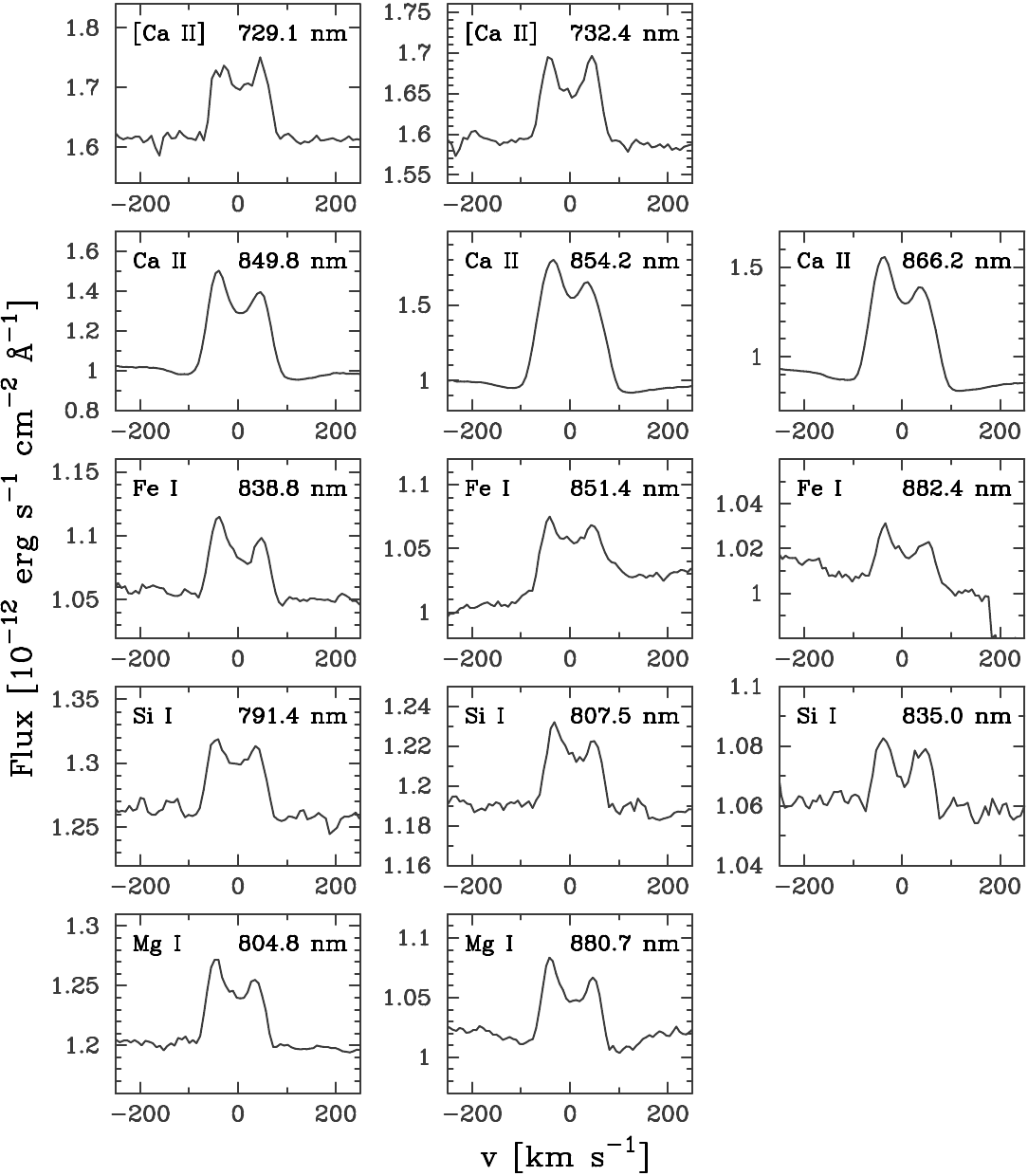}
   \caption{Detected emission lines in the XSHOOTER spectrum of HD\,152384. Velocity scales for all lines have been corrected for the radial velocity of $-10$ km~s$^{-1}$ derived from the photospheric absorption lines.}
              \label{Fig4}
    \end{figure}
The XSHOOTER spectrum of HD\,152384 (Fig.~\ref{Fig1}) is dominated by strong absorption lines from H\,{\sc i} and the Ca\,{\sc ii} H (396.8 nm) and K (393.4 nm) lines. From a Gaussian fit to these absorption lines we derive a heliocentric radial velocity for the star of $-10$ $\pm$ 2 km~s$^{-1}$. We compare the spectrum of HD\,152384 with a template spectrum for a A0 star (HD\,207222), taken from the XSHOOTER Spectral Library \citep{Gonneau2020}. The stellar absorption lines seen in this spectrum are a near-perfect match to the absorption lines seen in HD\,152384 (Fig.~\ref{Fig1}), confirming its previous spectral classification of A0\,V \citep{Houck1982}. We also derive a value for the stellar $v \sin i$ of 116 $\pm$ 20 km~s$^{-1}$ from the FWHM of the Mg\,{\sc ii} 448.1 nm line (seen in absorption) using the method of \cite{Royer2002}, correcting for the lower spectral resolution of XSHOOTER compared to the instrument used by Royer et~al.

   Using optical photometry retrieved from the SIMBAD database, as well as ultraviolet, near- and mid-IR data retrieved from the TD1 \citep{Thompson1978}, 2MASS \citep{Cutri2003} and AllWISE \citep{Cutri2013} catalogues, we construct the Spectral Energy Distribution of HD\,152384 (Fig.~\ref{Fig2}). As is shown in this Figure, the optical and near-IR photometry of HD\,152384 can be fit well by a mildly-extincted ($A_V$ = 0\fm09) model for the photosphere of a A0\,V star ($T_{\rm eff}$ = 9500~K), whereas a moderate ($L_{\rm IR}/L_\star$ = 0.05) excess emission above the expected photospheric levels can be seen longward of 2 microns. Using the integrated stellar flux from the fitted \cite{Kurucz1991} model for the stellar photosphere, as well as a distance of 128.3 $\pm$ 0.6 pc derived from the Gaia EDR3 parallax \citep{Gaia2020}, we derive a stellar radius of 1.9 $\pm$ 0.2 R$_\odot$ and stellar luminosity of 27.4 $\pm$ 2.1 L$_\odot$ for HD\,152384. Using the solar abundance pre-main sequence evolutionary tracks of \cite{Bressan2012} this combination of stellar temperature and luminosity corresponds to a 2.1 $\pm$ 0.1~M$_\odot$ star located close to the Zero-Age Main Sequence, indicating an age larger than 4.5 $\times$ 10$^6$ years. This lower limit to the stellar age of HD\,152384 is compatible with membership of the Sco-OB2 association \citep[e.g.][]{Preibisch2008}. Our derived stellar parameters are summarized in Table~\ref{Table1}.

   An inspection of archival
   %{\it Transiting Exoplanet Survey Satellite}
   TESS \citep[][]{Ricker2015} data for HD\,152384 show the star to be variable at the 0.1\% level, with a strict periodicity (Fig.~\ref{Fig3}). We estimate the period by fitting a Gaussian Process (GP) with a rotation kernel \citep[][eq. 56]{Foreman-Mackey2017}, which takes into account the presence of correlated noise in the TESS data. We find a period of $8.286 \pm 0.006$ hours, which we identify with the stellar rotation period. In addition to the regular brightness variations which we attribute to stellar rotation moving a different pattern of starspots in and out of our line of sight, we also note that a small (0.2\%) flare is present in the TESS data around 20.5 days, which could be due to further low-level stellar activity. 
Adopting a stellar radius of 1.9 R$_\odot$, the rotation period found above corresponds to a rotation velocity of $\sim$280 km~s$^{-1}$, or 62\% of the stellar break-up velocity of 460 km~s$^{-1}$. Although relatively fast for a main-sequence star, this rotation velocity is in-line with those found for other young early A-type stars \citep[][]{Wolff2004,Zorec2012}, providing further support for the classification of HD\,152384 as a young star. With the $v \sin i$ value derived from the Mg\,{\sc ii} 448.1 nm line, we also use the rotation velocity to derive a value for the stellar inclination of $i$ = 24 $\pm$ 4$^\circ$. 
\begin{table}
  \caption[]{Detected emission lines in the spectrum of HD\,152384.}
  \label{Table2}
 \vspace*{-0.4cm}
\begin{center}
\begin{tabular}{clc}
\hline\noalign{\smallskip}
$\lambda_{\rm lab.}$ [nm] & Ident. & EW [\AA]\\
\noalign{\smallskip}
\hline\noalign{\smallskip}
729.15  &  [Ca\,{\sc ii}]  & $-$0.20\\
732.39  &  [Ca\,{\sc ii}]  & $-$0.18\\
    
849.81  &  Ca\,{\sc ii}    & $-$1.60\\
854.21  &  Ca\,{\sc ii}    & $-$2.87\\
866.21  &  Ca\,{\sc ii}    & $-$2.44\\
    
838.78  &  Fe\,{\sc i}     & $-$0.14\\
851.41  &  Fe\,{\sc i}     & $-$0.19\\
882.41  &  Fe\,{\sc i}     & $-$0.10\\
    
804.78  &  Mg\,{\sc i}     & $-$0.15\\
880.67  &  Mg\,{\sc i}     & $-$0.14\\
    
791.38  &  Si\,{\sc i}     & $-$0.12\\
807.46  &  Si\,{\sc i}     & $-$0.09\\
835.04  &  Si\,{\sc i}     & $-$0.05\\
\noalign{\smallskip}
\hline
\end{tabular}
\end{center}
\end{table}

\section{Circumstellar Material}
There are several indications for the presence of circumstellar material in HD\,152384. The infrared excess emission seen longward of 2~$\mu$m in the SED (Fig.~\ref{Fig2}) suggests the presence of warm dust in the vicinity of the star. In addition, the XSHOOTER spectrum displays a number of emission lines superimposed on the photospheric absorption spectrum. We identify these lines with emission from Ca\,{\sc ii}, Si\,{\sc i}, Mg\,{\sc i} and Fe\,{\sc i} (Table~\ref{Table2}). All of the detected emission lines are centered on the stellar radial velocity, and thus appear to be clearly associated with HD\,152384. The emission lines display a clear double-peaked profile, with a peak-to-peak separation of around 70 km~s$^{-1}$ (Fig.~\ref{Fig4}).

The observed emission line profiles are reminiscent of the line profiles exhibited by the rotating gaseous disks seen around many young stars \citep[e.g.][]{vandenAncker2004}. These lines cannot be due to stellar activity, as (1) both the spectral type of A0\,V and the low level of variability seen in the TESS data suggest low levels of stellar activity, (2) stellar activity would result in emission components in the Ca\,{\sc ii} H and K lines \citep[e.g.][]{Vaughan1978}, which are not seen in our data and (3) stellar activity would not give the double-peaked emission line profiles seen in the XSHOOTER spectrum. The lines profiles are also unlikely to be solely due to a stellar wind as they do not exhibit a P Cygni like shape. We thus conclude that an origin in a circumstellar disk is the only plausible explanation for the presence of the emission lines seen in the XSHOOTER spectrum.

With the exception of the [Ca\,{\sc ii}] lines, which appear to be more symmetric, all detected emission lines display a distinct asymmetry, with the blue peak slightly higher than the red peak. This slight asymmetry could be a reflection of an asymmetric distribution of material within the disk, or be due to a wind from an optically thick disk \citep[e.g.][]{Ballabio2020,Pascucci2020}. Alternatively, there could be some additional extinction in our line-of-sight towards the part of the disk which is moving away from us. Higher spectral resolution data are needed to distinguish between these possibilities.

\begin{figure}
   \centering
   \includegraphics[angle=-90, width=8.0cm]{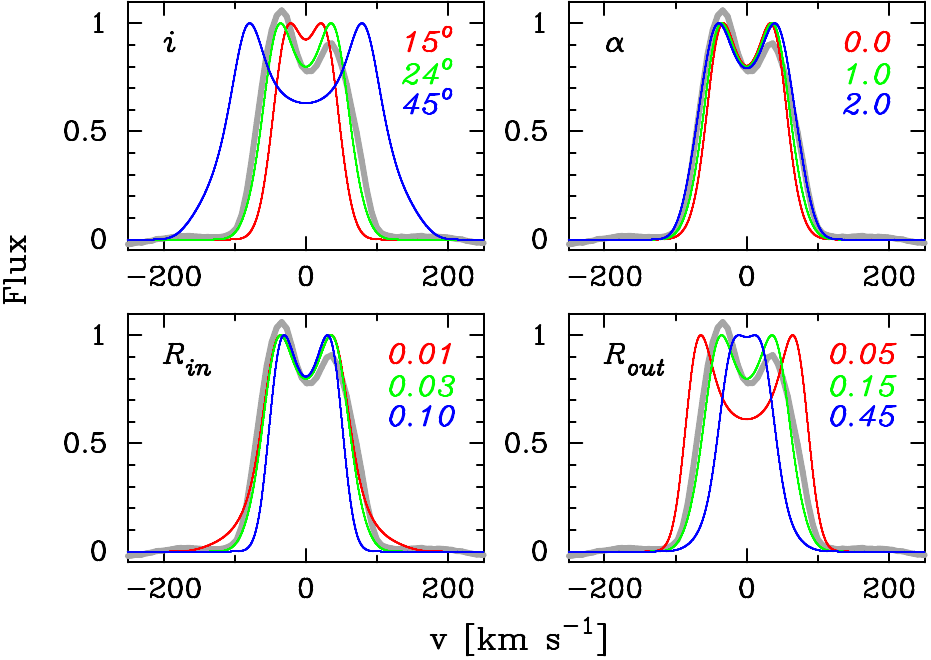}
   \caption{Simulated emission lines from a flat circumstellar disk using the toy disk model. The grey lines show the continuum-subtracted average Ca\,{\sc ii} line detected in HD\,152384, whereas the green lines show the model with the parameters $i = 24^\circ$ (corresponding to the stellar inclination), $\alpha = 1.0$, $R_{\rm in} = 0.03$ au, $R_{\rm out}$ = 0.15 au (corresponding to the best fit parameters). In each panel the red and blue lines illustrate the effect of varying one model parameter (top left: inclination, top right: $\alpha$, bottom left: disk inner radius in au, bottom right: disk outer radius in au).}
              \label{Fig5}
    \end{figure}

Following the approach in \cite{Carmona2011} we model the detected emission lines using a toy model for gas in a Keplerian orbit around HD\,152384 assuming a flat disk, with inner radius $R_{\rm in}$ and outer radius $R_{\rm out}$. The intensity of the emission decreases as $I(R) \propto (R/R_{\rm in})^{-\alpha}$, with $R$ the radial distance from the star. The total intensity of the model is computed as a function of velocity for a given inclination $i$ and convolved with a Gaussian with width corresponding to the XSHOOTER spectral resolution in the visual arm. Best fit parameters were selected by comparing a grid of simulated spectra with an average of the continuum-subtracted Ca\,{\sc ii} emission lines using a $\chi^2$ minimisation procedure. Examples of the resulting simulated spectra are shown in Fig.~\ref{Fig5}.
   
Assuming that the disk has the same inclination as derived for the stellar photosphere (24 $\pm$ 4$^\circ$), the emission from the disk needs to originate close to the star ($R_{\rm out}$ < 0.3 au) to be able to reproduce the double-peaked emission line profiles seen in the spectra. However, we note that in our model the disk inclination is to some extent degenerate with the disk outer radius, so the emanating region would move correspondingly further out if the disk is seen under higher inclination than the value derived from the stellar photosphere. The lack of high-velocity wings in the observed line profile may indicate the presence of a small inner gap, which can be reproduced in our model by setting $R_{\rm in}$ to a value of around 0.03 au. Varying $\alpha$, the exponent of the intensity distribution, has only a small effect on the width of the computed line profile. It is therefore poorly constrained by our observations.

What makes the emission-line spectrum of HD\,152384 highly unusual is the complete absence of Hydrogen and Helium emission (c.f. Fig.~\ref{Fig1}), suggesting that the disk is highly depleted in volatile elements. To the best of our knowledge this spectrum with strong emission lines from refractory elements, but completely lacking emission from volatile species is unique within the group of young stars. However, similar spectra are seen in rare instances in a completely different class of stars: metal-polluted white dwarfs \citep[][]{Gansicke2006,Wilson2014,Manser2016,Dennihy2020,Melis2020,GentileFusillo2021}. Similar to HD\,152384, these rare emission line systems also exhibit excess infrared emission in their SED caused by the presence of circumstellar dust. In the case of these white dwarfs the dust disk is the result of the tidal disruption of rocky planetesimals \citep{Veras2015,Malamud2020}, and the double-peaked emission lines in their spectrum is thought to originate from a gaseous component in these circumstellar debris disks. The exact mechanism generating the gas is still a matter of debate and  the proposed theories include: dust sublimation at the inner edge of the debris disk followed by radial spreading \citep{Metzger2012}, collisional cascades crushing the debris into gas \citep{Kenyon2017}, and the presence of a dense planetesimal orbiting within the disc and disrupting the dust \citep{Manser2019}.

\section{Discussion and Conclusions}
HD\,152384 appears to be unique within the group of young intermediate-mass stars with disks. Its spectrum shows double-peaked Ca, Si, Mg and Fe emission lines, but no Hydrogen or Helium emission, indicative of a disk which, although rich in refractory elements, is highly depleted in volatile materials. The slightly higher blue peak seen in most lines could be due to a slight asymmetry in the disk, could indicate the presence of a disk wind, or that there could be some additional extinction towards the redshifted disk material. Assuming that star and disk are seen under the same inclination angle of $\sim24^\circ$, the width and peak-to-peak separation of the detected double-peaked emission lines would suggest an origin of these lines in a region close (radius $<$ 0.3 au) to the star.

These properties of HD\,152384, although unique for a young star, bear a striking resemblance to a rare subset of metal-polluted White Dwarfs, in which the presence of an emission-line spectrum composed solely of refractory elements is commonly explained by the destruction of rocky planets \citep[][]{Bonsor2017}. We speculate that similar processes could occur around a young star with a newly formed exoplanetary system, in which the disruption caused by planetary migration could cause collisions between rocky planets. We note that the compact ($<$ 0.3 au) disk around HD\,152384 is smaller than the disks seen in most Herbig Ae/Be stars \citep[][and references therein]{Kraus2015}, but its location is similar to rocky planets in our own solar system. In addition, simulations of tidal disruption of giant planets due to interactions with their host star show that remnants of exomoons can end up in orbits that are similar to the disk seen in HD\,152384 \citep[][]{Martinez2019}. Interestingly, the relatively low fractional infrared luminosity ($L_{\rm IR}/L_\star$ = 0.05) places HD\,152384 within the group of extreme debris disks defined recently by \cite{Moor2021}, for which large amounts of warm dust may stem from recent giant impacts between planetary embryos during the final phases of terrestrial planet growth. Although cool molecular gas has been found at relatively large distances in a number of these objects \citep{Moor2017,Marino2020}, our analysis on HD\,152384 constitutes the first detection of gas solely consisting of refractory elements around such an object, leading further support for the hypothesis of the destruction of planetary embryos. 

Signatures of selective accretion of circumstellar gas have also been found in the abundance patterns seen in the photospheres of $\lambda$ Boo stars and Herbig stars \citep{Venn1990,Folsom2012,Kama2015,Jermyn2018}, and may be due to Jupiter-like planets blocking the accretion of part of the dust. Several systems of similar age as HD\,152384 such as $\beta$ Pic also display variable absorption features in their spectrum which are interpreted as due to infalling evaporating comets \citep[e.g.][]{Ferlet1987,Beust1991,Kiefer2014,Vidal-Madjar2017,Rebollido2020}, demonstrating the dynamic nature of newly formed planetary systems. Collisions are also believed to have played an important role in shaping our own solar system, and may have led to the formation of our Earth's moon \citep[e.g.][]{Canup2001}. We postulate that HD\,152384 may currently be experiencing a similarly dynamic phase in its evolution. One prediction of our hypothesis of disruption by planetary migration is that HD\,152384 should possess one or more giant planets at radii larger than a few tens of an au which would be responsible for the disruption of the inner rocky planets and the creation of the circumstellar disk reported here. High contrast imagers on the next generation of Extremely Large Telescopes will likely be able to detect these.

\begin{acknowledgements}
We thank the ESO staff at Paranal and in Garching for their support during the preparation and execution of the observations. We would also like to thank the referee for providing insightful and constructive comments.
\end{acknowledgements}

 \bibliographystyle{apa}
 \bibliography{bibliography}

\end{document}